\documentclass[
    onecolumn,
	prd,
	amssymb,
	preprintnumbers,superscriptaddress,
	nofootinbib,notitlepage]{revtex4-2}
\pdfoutput=1
\usepackage{paralist}
\usepackage{graphicx}
\usepackage{enumitem}
\usepackage{latexsym}
\usepackage{amsfonts}
\usepackage{amssymb}
\usepackage{xcolor}
\usepackage[export]{adjustbox}
\usepackage{amsmath}
\usepackage[thinlines]{easytable}
\usepackage{slashed}
\usepackage{dcolumn}
\usepackage{verbatim}
\usepackage{float}
\usepackage{multirow}
\usepackage{xspace}
\usepackage[normalem]{ulem}
\usepackage[
pdfauthor={Jeremy Sakstein}]{hyperref}
\usepackage{tabularx}
\usepackage{lettrine}
\usepackage{setspace}
\usepackage{float}
\input Zallman.fd

\LettrineTextFont{\itshape}

\renewcommand{\tilde}{\widetilde} 

\newcommand{\beq}{\begin{equation}}
\newcommand{\eeq}{\end{equation}}
\newcommand{\bea}{\begin{eqnarray}}
\newcommand{\eea}{\end{eqnarray}}

\newcommand{\refjnl}[1]{{\rm#1}}

\def\mnras{\refjnl{MNRAS}}             
\def\nat{\refjnl{Nature}}              

\usepackage{pdfbase}[2017/03/16]
\usepackage{xparse,ocgbase}
\usepackage{xcolor,calc}
\usepackage{tikzpagenodes,linegoal}
\usetikzlibrary{calc}
\usepackage{tcolorbox}

\ExplSyntaxOn
\let\tpPdfLink\pbs_pdflink:nn
\let\tpPdfAnnot\pbs_pdfannot:nnnn\let\tpPdfLastAnn\pbs_pdflastann:
\let\tpAppendToFields\pbs_appendtofields:n
\def\tpPdfXform{\pbs_pdfxform:nnnnn{1}{1}{}{}}
\let\tpPdfLastXform\pbs_pdflastxform:
\let\cListSet\clist_set:Nn\let\cListItem\clist_item:Nn
\ExplSyntaxOff

\usepackage{pdfbase}[2017/03/16]
\usepackage{xparse,ocgbase}
\usepackage{xcolor,calc}
\usepackage{tikzpagenodes,linegoal}
\usetikzlibrary{calc}
\usepackage{tcolorbox}

\ExplSyntaxOn
\let\tpPdfLink\pbs_pdflink:nn
\let\tpPdfAnnot\pbs_pdfannot:nnnn\let\tpPdfLastAnn\pbs_pdflastann:
\let\tpAppendToFields\pbs_appendtofields:n
\def\tpPdfXform{\pbs_pdfxform:nnnnn{1}{1}{}{}}
\let\tpPdfLastXform\pbs_pdflastxform:
\let\cListSet\clist_set:Nn\let\cListItem\clist_item:Nn
\ExplSyntaxOff

\makeatletter
\NewDocumentCommand{\tooltip}{%
  ssssO{\ifdefined\@linkcolor\@linkcolor\else blue\fi}mO{yellow!20}mO{0pt,0pt}%
}{{%
  \leavevmode%
  \IfBooleanT{#2}{%
    \ocgbase@new@ocg{tipOCG.\thetcnt}{%
      /Print<</PrintState/OFF>>/Export<</ExportState/OFF>>%
    }{false}%
    \xdef\tpTipOcg{\ocgbase@last@ocg}%
    \ocgbase@add@ocg@to@radiobtn@grp{tool@tips}{\ocgbase@last@ocg}%
  }%
  \tpPdfLink{%
    \IfBooleanTF{#4}{%
      /Subtype/Link/Border[0 0 0]/A <</S/SetOCGState/State [/Toggle \tpTipOcg]>>
    }{%
      /Subtype/Screen%
      /AA<<%
        \IfBooleanTF{#3}{%
          /E<</S/SetOCGState/State [/Toggle \tpTipOcg]>>%
        }{%
          \IfBooleanTF{#2}{%
            /E<</S/SetOCGState/State [/ON \tpTipOcg]>>%
            /X<</S/SetOCGState/State [/OFF \tpTipOcg]>>%
          }{
            \IfBooleanTF{#1}{%
              /E<</S/JavaScript/JS(%
                var fd=this.getField('tip.\thetcnt');%
                if(typeof(click\thetcnt)=='undefined'){%
                  var click\thetcnt=false;%
                  var fdor\thetcnt=fd.rect;var dragging\thetcnt=false;%
                }%
                if(fd.display==display.hidden){%
                  fd.delay=true;fd.display=display.visible;fd.delay=false;%
                }else{%
                  if(!click\thetcnt&&!dragging\thetcnt){fd.display=display.hidden;}%
                  if(!dragging\thetcnt){click\thetcnt=false;}%
                }%
                this.dirty=false;%
              )>>%
            }{%
              /E<</S/JavaScript/JS(%
                var fd=this.getField('tip.\thetcnt');%
                if(typeof(click\thetcnt)=='undefined'){%
                  var click\thetcnt=false;%
                  var fdor\thetcnt=fd.rect;var dragging\thetcnt=false;%
                }%
                if(fd.display==display.hidden){%
                  fd.delay=true;fd.display=display.visible;fd.delay=false;%
                }%
               this.dirty=false;%
              )>>%
              /X<</S/JavaScript/JS(%
                if(!click\thetcnt&&!dragging\thetcnt){fd.display=display.hidden;}%
                if(!dragging\thetcnt){click\thetcnt=false;}%
                this.dirty=false;%
              )>>%
            }%
            /U<</S/JavaScript/JS(click\thetcnt=true;this.dirty=false;)>>%
            /PC<</S/JavaScript/JS (%
              var fd=this.getField('tip.\thetcnt');%
              try{fd.rect=fdor\thetcnt;}catch(e){}%
              fd.display=display.hidden;this.dirty=false;%
            )>>%
            /PO<</S/JavaScript/JS(this.dirty=false;)>>%
          }%
        }%
      >>%
    }%
  }{{\color{#5}#6}}%
  \sbox\tiptext{%
    \IfBooleanT{#2}{%
      \ocgbase@oc@bdc{\tpTipOcg}\ocgbase@open@stack@push{\tpTipOcg}}%
    \tcbox[colframe=black,colback=#7,size=fbox,arc=1ex,sharp corners=southwest]{#8}%
    \IfBooleanT{#2}{\ocgbase@oc@emc\ocgbase@open@stack@pop\tpNull}%
  }%
  \cListSet\tpOffsets{#9}%
  \edef\twd{\the\wd\tiptext}%
  \edef\tht{\the\ht\tiptext}%
  \edef\tdp{\the\dp\tiptext}%
  \tipshift=0pt%
  \IfBooleanTF{#2}{%
    \setlength\whatsleft{\linegoal}%
  }{%
    \measureremainder{\whatsleft}%
  }%
  \ifdim\whatsleft<\dimexpr\twd+\cListItem\tpOffsets{1}\relax%
    \setlength\tipshift{\whatsleft-\twd-\cListItem\tpOffsets{1}}\fi%
  \IfBooleanF{#2}{\tpPdfXform{\tiptext}}%
  \raisebox{\heightof{#6}+\tdp+\cListItem\tpOffsets{2}}[0pt][0pt]{%
    \makebox[0pt][l]{\hspace{\dimexpr\tipshift+\cListItem\tpOffsets{1}\relax}%
    \IfBooleanTF{#2}{\usebox{\tiptext}}{%
      \tpPdfAnnot{\twd}{\tht}{\tdp}{%
        /Subtype/Widget/FT/Btn/T (tip.\thetcnt)%
        /AP<</N \tpPdfLastXform>>%
        /MK<</TP 1/I \tpPdfLastXform/IF<</S/A/FB true/A [0.0 0.0]>>>>%
        /Ff 65536/F 3%
        /AA <<%
          /U <<%
            /S/JavaScript/JS(%
              var fd=event.target;%
              var mX=this.mouseX;var mY=this.mouseY;%
              var drag=function(){%
                var nX=this.mouseX;var nY=this.mouseY;%
                var dX=nX-mX;var dY=nY-mY;%
                var fdr=fd.rect;%
                fdr[0]+=dX;fdr[1]+=dY;fdr[2]+=dX;fdr[3]+=dY;%
                fd.rect=fdr;mX=nX;mY=nY;%
              };%
              if(!dragging\thetcnt){%
                dragging\thetcnt=true;Int=app.setInterval("drag()",1);%
              }%
              else{app.clearInterval(Int);dragging\thetcnt=false;}%
              this.dirty=false;%
            )%
          >>%
        >>%
      }%
      \tpAppendToFields{\tpPdfLastAnn}%
    }%
  }}%
  \stepcounter{tcnt}%
}}
\makeatother
\newsavebox\tiptext\newcounter{tcnt}
\newlength{\whatsleft}\newlength{\tipshift}
\newcommand{\measureremainder}[1]{%
  \begin{tikzpicture}[overlay,remember picture]
    \path let \p0 = (0,0), \p1 = (current page.east) in
      [/utils/exec={\pgfmathsetlength#1{\x1-\x0}\global#1=#1}];
  \end{tikzpicture}%
}

\newcommand{\dd}{\mathrm{d}}

\newcommand{\msun}{{\rm M}_\odot}

\DeclareRobustCommand{\okina}{%
  \raisebox{\dimexpr\fontcharht\font`A-\height}{%
    \scalebox{0.8}{`}%
  }%
}

\allowdisplaybreaks

\begin{document}

\title{Slowly Rotating Neutron Stars in Aether Scalar-Tensor Theory}

\author{Christopher Reyes}
    \email{cmreyes3@hawaii.edu}
    \affiliation{Department of Physics \& Astronomy, University of Hawai\okina i, Watanabe Hall, 2505 Correa Road, Honolulu, HI, 96822, USA}
    \author{Jeremy Sakstein} \email{sakstein@hawaii.edu}
\affiliation{Department of Physics \& Astronomy, University of Hawai\okina i, Watanabe Hall, 2505 Correa Road, Honolulu, HI, 96822, USA}

\date{\today}

\begin{abstract}
Aether Scalar-Tensor theory is a relativistic alternative gravity model that behaves like cold dark matter on cosmological scales while predicting the MOND force-law in astrophysical systems.~The theory correctly predicts the cosmic microwave background and linear matter power spectra, and the mass discrepancies observed across the Universe.~We derive and solve the equations governing neutron stars in Aether Scalar Tensor theory at first-order in slow rotation, finding that the theory predicts approximate universal relations between the moment of inertia and the compactness ($I$--$C$ relations) that differ from their general relativity counterparts.~These relations may enable tests of Aether Scalar-Tensor theory using X-ray observations of pulsars and gravitational wave observations of binary neutron star mergers.~
\end{abstract}

\maketitle

\section{Introduction}
\label{sec:intro}

It has long been known that General relativity (GR) fails to reproduce the observed phenomenology of the Universe on astrophysical and cosmological scales when applied to luminous matter.~Prominent examples of such systems include galaxies, where the shape of rotation curves and the lensing of light is not predicted by optical observations of their constituent stars and gas;~and cosmology, where the power spectra of the cosmic microwave background (CMB) and large scale structure (LSS), cannot be explained by baryons.~Further examples are discussed in \cite{Famaey:2011kh}.~Hypotheses for explaining these systems' behavior generally fall into two categories:~either the Universe contains non-luminous dark matter (DM) that contributes to these objects' gravitational potential but not their luminosity, or GR is modified at accelerations smaller than the scale $a_0\sim 1.2\times10^{-10}$ m/s$^2$, the regime where GR applied to visible matter is unable to predict our observations.~The quintessential paradigm for modified gravity (MG) is modified Newtonian dynamics (MOND) \cite{Milgrom:1983ca,Bekenstein:1984tv,Milgrom:2009ee,Famaey:2011kh,Banik:2021woo}, which modifies Newton's law of gravitation when $a<a_0$.

Both proposed explanations have successes and challenges.~The DM paradigm correctly predicts observations on large scales where cosmological perturbations are linear such as the CMB and LSS but it has difficulty reproducing small-scale phenomena from first principles.~Examples include the Tully-Fisher \cite{Tully:1977fu} and Faber-Jackson \cite{Faber:1976sn} relations, the cored density profiles of some galaxies \cite{Ostriker:2003qj,DelPopolo:2021bom}, the masses and phase-space distribution of Milky Way satellite galaxies \cite{2011MNRAS.415L..40B,2012MNRAS.423.1109P,Pawlowski:2013cae,2014Natur.511..563I,Ibata:2013rh}, and the dynamics of tidal dwarf galaxies \cite{Gentile:2007gp,Kroupa:2012qj,Kroupa:2014ria}.~The DM paradigm invokes highly-uncertain baryonic physics to explain these phenomena.~In contrast, MG models correctly predict the dynamics of astrophysical objects but have difficulty explaining cosmological observations.~Many of the proposed theories \cite{Bekenstein:1988zy,Sanders:1996wk,Bekenstein:2004ne,Moffat:2005si,Navarro:2005ux,Zlosnik:2006zu,Sanders:2005vd,Milgrom:2009gv,Babichev:2011kq,Deffayet:2011sk,Blanchet:2011wv,Sanders:2011wa,Mendoza:2012hu,Woodard:2014wia,Khoury:2014tka,Blanchet:2015sra,Hossenfelder:2017eoh,Burrage:2018zuj,Milgrom:2019rtd,DAmbrosio:2020nev,Kading:2023hdb} are unable to match our observations while others are now excluded because they predict large differences between the speed of light and gravitational waves, a possibility that is excluded at the $10^{-16}$ level by GW170817 \cite{LIGOScientific:2017vwq,Sakstein:2017xjx,Baker:2017hug,Creminelli:2017sry,Ezquiaga:2017ekz}.

Recently, a hybrid model that contains elements of both DM and MG has been proposed:~Aether Scalar-Tensor theory (AeST) \cite{Skordis:2020eui}.~AeST predicts the MOND force-law at accelerations smaller than $a_0$ so  explains the astrophysical phenomena above from first principles.~On cosmological scales, the theory behaves as a ghost condensate \cite{Arkani-Hamed:2003pdi,Furukawa:2010gr} contributing a pressureless fluid with zero sound speed  to the universe that mimics cold dark matter.~Thus, the theory correctly predicts LSS and the CMB.~The theory propagates six healthy degrees of freedom \cite{Bataki:2023uuy} and the GW170817 bound is satisfied because the spin-2 degrees of freedom propagating luminally.~These promising features have spurred an intense research effort studying AeST~\cite{Mistele:2021qvz,Kashfi:2022dyb,Bernardo:2022acn,Mistele:2023paq,Mistele:2023fwd,Verwayen:2023sds,Durakovic:2023out,Llinares:2023lky,Tian:2023gjt,Rosa:2023qun,Rosa:2024fwc,Hsu:2024ftc,Skordis:2024wlo}.

The theory is viable alternative to GR, so testing it against observations is paramount.~In this work, we take steps towards probing AeST using X-ray and gravitational wave observations of neutron stars (NSs).~Previously, we have derived the properties of spherically-symmetries NSs in AeST \cite{Reyes:2024oha}.~In that work we established the existence of neutron stars in AeST and showed that their properties are consistent with observations.~This is an important first step, but the unknown nature of the equation of state (EOS) of nuclear matter prevents tests of the theory using static NSs.~This uncertainty can be mitigated using \textit{universal relations} between the NSs compactness $C=GM/R$, its moment of inertia $I$, and its mass moments e.g., the tidal Love number and quadrupolar deformability \cite{Yagi:2013bca,Yagi:2013awa,Breu:2016ufb,Yagi:2016bkt,Doneva:2017jop}.~These relations are independent of the EOS and therefore measuring them can provide bounds on MG theories e.g., \cite{Silva:2020acr,Saffer:2021gak}.~

In this work, we derive two relations between the moment of inertia and compactness of NSs in AeST by deriving and solving the equations governing these objects at first-order in slow rotation.~We find that universal relations exist in AeST, and that they differ from the GR predictions.~We provide fitting formulae for these relations as a function of the model parameters.~Our predictions may help to enable NS tests of AeST, and lay the foundation for finding further relations between higher moments e.g., the I-Love-Q relations.

This work is organized as follows.~We  review the salient features of AeST in section~\ref{sec:Aest_review}.~The equations governing slowly-rotating NSs are derived and solved in section~\ref{sec:TOV}, where we also present out modified $I$--$C$ relations.~We conclude in section~\ref{sec:conclusions}.

\section{Review of Aether Scalar-Tensor Theory}
\label{sec:Aest_review}

Aether Scalar-Tensor theory is described by the following action \cite{Skordis:2020eui}:
\begin{align}
S&=\int\mathrm{d}^4x\frac{\sqrt{-g}}{16\pi G}\left[R-\frac{K_B}{2}{F}^{\mu\nu}{F}_{\mu\nu}-\lambda(A_\mu A^\mu +1)+(2-K_B)(2J^\mu\nabla_\mu\phi-\mathcal{Y})-\mathcal{F}(\mathcal{Y},\mathcal{Q})\vphantom{\frac{K_B}{2}{F}^{\mu\nu}{F}_{\mu\nu}}\right] +S_{\rm m}[g],\label{eq:action}
\end{align}
where $g_{\mu\nu}$ is the space-time metric, $\phi$ is a scalar, and \textit{the aether} $A^\mu$ is a unit-timelike vector $A_\mu A^{\mu}=-1$.~The other quantities appearing in the action are given by   $F_{\mu\nu}=2\nabla_{[\mu}A_{\nu]}$ with $\nabla_\mu$ the metric-compatible connection, $J_\mu=A^\nu\nabla_\nu A_\mu$, $\mathcal{Q}=A^\mu\nabla_\mu\phi$, $\mathcal{Y}=q^{\mu\nu}\nabla_\mu\phi\nabla_\nu\phi$ with $q_{\mu\nu}=g_{\mu\nu}+A_\mu A_\nu$.~The function $\mathcal{F}(\mathcal{Y},\mathcal{Q})$ is arbitrary and $K_B$ is a constant.~Finally, $\lambda$ is a Lagrange multiplier that imposes the unit-time-like constraint for $A_\mu$.~

The free function $\mathcal{F}(\mathcal{Y},\mathcal{Q})$ is chosen to ensure that the theory replicates the predictions of $\Lambda$CDM on cosmological scales and behaves like MOND in weak-field quasi-static situations.~In practice, this requires that functional form of $\mathcal{F}(\mathcal{Y},\mathcal{Q})$ differs in strong and weak gravitational backgrounds.~The characteristic acceleration of static neutron stars is far in excess of Milgrom’s constant $a_0=1.2\times10^{-10}$ m/s$^2$, $GM/R_{NS}^{2} \gg a_0$ with $R_{NS}\sim 10$ km and $M\sim 1.5 M_{\odot}$, placing these objects in the strong-field limit of the theory in which $\mathcal{F}(\mathcal{Y},\mathcal{Q})$ is taken as $(2-K_{B})\lambda_{\rm s}\mathcal{Y} + \mathcal{K}_2(\mathcal{Q}-\mathcal{Q}_0)^2$, with $\lambda_{\rm s}$ a free parameter that is related to the GR limit of the theory \cite{Skordis:2020eui,Skordis:2021mry,Verwayen:2023sds}.~This parameter is necessarily positive for the theory to be stable on Minkowski space \cite{Skordis:2021mry}.~The term proportional to $\mathcal{K}_2$ is responsible for ensuring compatibility with $\Lambda$CDM on cosmological scales, and  introduces a mass term $\mu$ for the metric potentials that must be taken to satisfy $\mu\sim {\rm Mpc}^{-1}$ to reproduce cosmological observables \cite{Skordis:2020eui}.~Since $\mu{R}_{\rm NS}\ll1$ this mass term can be neglected in comparison with the factor $(2-K_{\rm B})\lambda_{\rm s}\mathcal{Y}$.~Newton's constant as measured by solar system experiments is given by $G_N=(1+\lambda_{\rm s}^{-1})(1-K_B/2)^{-1}G$ \cite{Skordis:2020eui,Skordis:2021mry}.~This implies that $0<K_B<2$ to ensure that the graviton and vector have correct-sign kinetic terms.

The field equations obtained by varying the action with respect the metric, the scalar, and the vector are as follows.~The modified Einstein equation is
\begin{eqnarray}
    &&G_{\mu\nu}  =8\pi GT^{\rm AeST}_{\mu\nu}+ 8\pi G T_{\mu\nu}\label{eq:Modified},
\end{eqnarray}
where $T_{\mu\nu}$ is the  matter energy-momentum tensor and $T^{\rm AeST}_{\mu\nu}$ is the effective AeST energy momentum tensor given by: 
\begin{align}
 8\pi G T^{\rm AeST}_{\mu\nu} &=   K_BF_{\mu}^\alpha F_{\nu\alpha} - 
    (2-K_{B}) \{ 2J_{(\mu}\nabla_{\nu)}\phi - A_{\mu}A_{\nu}\nabla^{\alpha}\nabla_{\alpha}\phi \nonumber 
     + 2\left[A_{(\mu}\nabla_{\nu)}A_{\alpha} - A_{(\mu}\nabla_{|\alpha|}A_{\nu)}\right]\nabla^{\alpha}\phi\}   \nonumber \\ 
     &+ \mathcal{F}_{\mathcal{Q}}A_{(\mu}\nabla_{\nu)}\phi + (2-K_{B}+ \mathcal{F}_{\mathcal{Y}}) [ \nabla_{\mu}\phi\nabla_{\nu}\phi + 2\mathcal{Q}A_{(\mu}\nabla_{\nu)}\phi]  
      +\lambda A_{\mu}A_{\nu}\nonumber\\
      &-  \frac12g_{\mu\nu}\left(\frac{K_B}{2}{F}^{\mu\nu}{F}_{\mu\nu}+\lambda(A_\mu A^\mu +1)-(2-K_B)(2J^\mu\nabla_\mu\phi-\mathcal{Y})+\mathcal{F}(\mathcal{Y},\mathcal{Q})\vphantom{\frac{K_B}{2}{F}^{\mu\nu}{F}_{\mu\nu}}\right),
 \end{align}   
where $\mathcal{F}_{\mathcal{Y}}= \partial _{\mathcal{Y}}\mathcal{F}$,  $\mathcal{F}_{\mathcal{Q}}= \partial _{\mathcal{Q}}\mathcal{F}$.~The scalar's equation of motion is
\begin{align}
\nabla_{\mu}\mathcal{J}^{\mu} =0,\label{eq:scalar}
\end{align}
where $\mathcal{J}^{\mu}$ = $(2-K_{B})J^{\mu} -  (2-K_{B}+\mathcal{F}_{\mathcal{Y}})q^{\alpha\mu}\nabla_{\alpha}\phi - \mathcal{F}_{\mathcal{Q}}A^{\mu}/2$.~The vector's equation of motion is 
\begin{eqnarray}
&&K_{B}\nabla_{\nu}F^{\nu\mu} + (2-K_{B})\left[(\nabla^{\nu}A^{\mu})\nabla_{\nu}\phi - \nabla_{\nu}(A^{\nu}\nabla^{\mu}\phi) \right] 
 -\left[ (2-K_{B} + \mathcal{F}_{\mathcal{Y}}\mathcal{Q}) + \mathcal{F}_{\mathcal{Q}}/2\right]\nabla^{\mu}\phi - \lambda A^{\mu} = 0\label{eq:vector}.
\end{eqnarray}
Finally, the equation of motion for Lagrange multiplier $\lambda$ imposes the constraint
\begin{align}
    A^{\mu}A_{\mu} +1 =0.\label{eq:constrain}
\end{align}

\section{Slowly-Rotating Neutron Stars}
\label{sec:TOV}

The metric for a static spherically-symmetric slowly rotating NSs up to first order in angular momentum, can be written as~\cite{Hartle:1968si}
\begin{equation}
    \dd s^{2}= -e^{\alpha(r)}\dd t^{2}
     +e^{2\gamma(r)}\dd r^{2} +r^{2}
 \dd\theta^{2} + r^2\sin^{2}\theta\left(d\phi -\varepsilon\left[\Omega_{*}- \omega(r)P'_{1}(\cos\theta) \right]dt \right)  ^{2}
 \label{eq:metric}
\end{equation}
where, $\alpha$, $\gamma$ and $\omega$ are metric potentials that only depend on $r$, $\varepsilon$ is a bookkeeping parameter that tracks the order of the perturbations and $\Omega_{*}$ is the angular velocity of the NS, which is taken to be much smaller than the characteristic angular velocity of the star, $\Omega_{*}^{2} \ll GM/R^{3} $, where $M$ and $R$ are the mass and radius of the NS respectively. We only included the $l = 1$ odd perturbation, since our goal is to calculate the moment of inertia, which is sourced by $l = 1$ odd (axial) modes \cite{Yagi:2013awa,Breu:2016ufb,Yagi:2016bkt}.~We  write the scalar as $\phi=\phi(r)$.~This is not the most general ansatz we can write for the scalar and, in fact, a time dependence of the form $\phi(t,r)=qt+\varphi(r)$ is needed for compatibility with cosmology \cite{Skordis:2020eui} but, as  discussed in our previous work \cite{Reyes:2024oha}, since we are only interested in local behavior of gravity around neutron stars, where the length scale is much smaller than the size of the Universe, we do not expect this to have any appreciable effect on the neutron stars.~

To first order the scalar field is expanded as
\begin{equation}
    \phi = \phi_0(r) +\varepsilon \phi_1(\vec{r}),
\end{equation}
where $\phi_0(r)$ is the scalar field of the static, spherically symmetric neutron star, and $\phi_1(\vec{r})$—which takes the form $\phi_1(r)P_{l}(\cos\theta)$—is the field induced by slow rotation. However, since we are only including $l = 1$ odd perturbations and $\phi_1(r)P_{1}(\cos\theta)$ is even, we set $\phi_1(\vec{r}) = 0$. That is, no odd scalar perturbations are present.~

Turning to the vector, we write this as
\begin{equation}
    A_{\mu}\dd x^{\mu}= e^{\frac{\alpha(r)}{2}}\dd t  + \varepsilon B(r)\sin(\theta)^{2}\dd\phi,  \label{eq:vectorAnt}
\end{equation}
where we have included only $l=1$ odd perturbations.~In general, the vector could have a radial component, but to ensure consistency with our previous work \cite{Reyes:2024oha} we chose the vector to be aligned in the time direction at zeroth-order in slow rotation, which greatly simplifies the equations.~

Finally, we treat the neutron star as a perfect fluid with energy-momentum tensor given by
\begin{equation}
    T_{\mu\nu} = \left( \rho +P  \right)u_{\mu}u_{\nu} 
    + Pg_{\mu\nu},\label{eq:EM}
\end{equation}
where $\rho$ and $P$ are the energy density and pressure of the fluid with four-velocity
\begin{equation}
u^{\mu}\partial_{\mu} = e^{-\frac{\alpha(r)}{2}}\partial_{t} + \Omega_{*}e^{-\frac{\alpha(r)}{2}} \partial_{\phi}.~\label{eq:vel}
\end{equation}

We now proceed to solve the coupled Einstein, vector, and scalar equations up to first-order in slow rotation, $\varepsilon$.~The zeroth-order equations were derived in our previous publication \cite{Reyes:2024oha} and are presented because their are solutions are necessary inputs for solving the first-order equations.

\subsection{Background Equations and Static Spherically-Symmetric Neutron Stars}

\begin{figure}[ht]
    \centering
\includegraphics[width=0.9\textwidth]{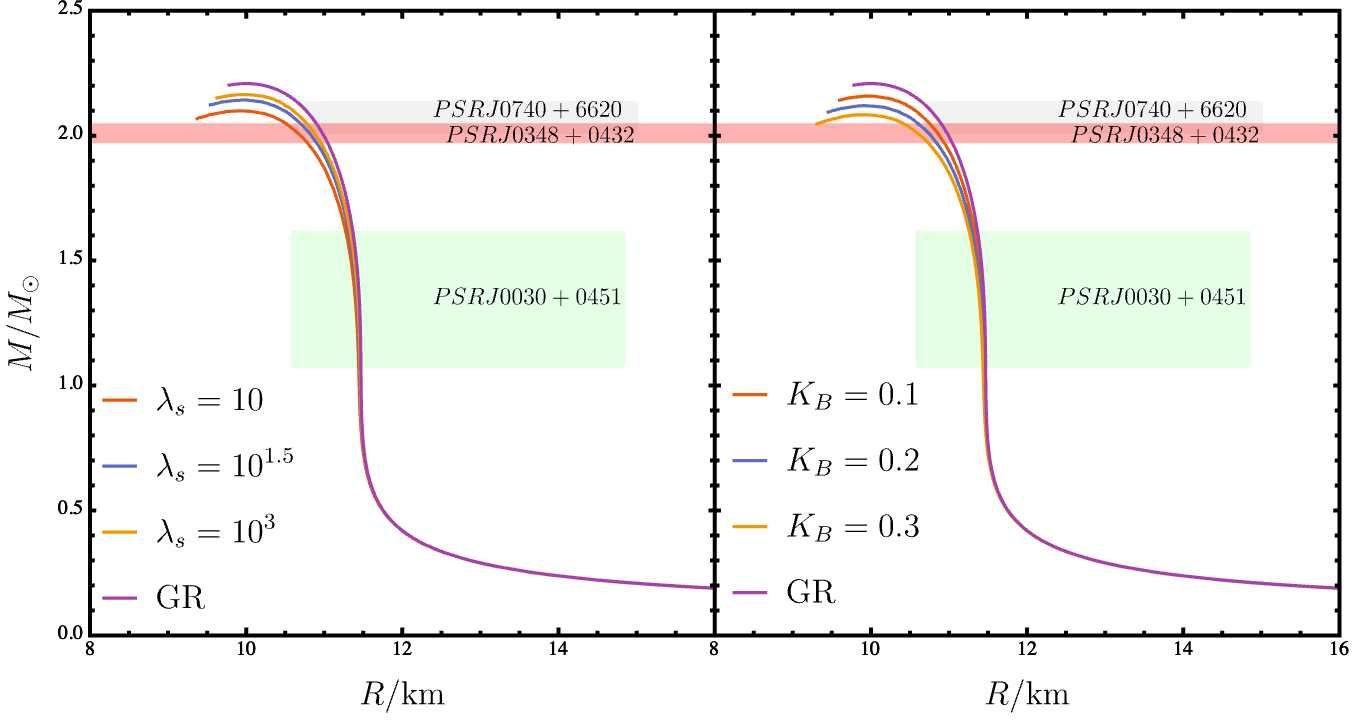}
    \caption{Mass-radius relations for the AeST parameters indicated in the figure assuming the APR EOS.~The left panel shows the effect of varying $\lambda_{s}$ at fixed $K_{B}=0.1$, and the right panel shows the effect of varying $K_{B}$ at fixed $\lambda_{s}=10^2$.}
    \label{fig:MR_APR}
\end{figure}

To zeroth-order $(\varepsilon^0)$ in slow rotation, the stellar structure equations are the following (see \cite{Reyes:2024oha}, appendix A for a detailed derivation):
\begin{equation}
    \frac{dP}{dr} = - \frac12[P(r)+\rho(r)]\alpha^{\prime}(r),\label{eq:dp1}
\end{equation}
\begin{equation}
    \alpha ''(r)+\frac{\Upsilon(r;K_B,\lambda_{s})}{8 (\lambda_{\rm s} +1) r \left(\lambda_{\rm s} +1-4 \pi  G_{N} \lambda_{\rm s}  (K_B-2) r^2 P\right)}=0\label{eq:alpOD1}
\end{equation}
where $\Upsilon(r;K_B,\lambda_{s})$ is given in \cite{Reyes:2024oha}.~Near the center of the NS up to $\mathcal{O}(r^2)$ the solutions of \eqref{eq:dp1} and \eqref{eq:alpOD1} behave as 
\begin{align}
    \alpha(r_{i}) &= \alpha_{c} + \frac{8\pi G_N(3p_{c}+\rho_{c})}{3}r_{i}^{2}
    + \mathcal{O}(r_{i}^{4})\label{eq:alpcen}\\
    \rho(r_{i}) &= \rho_{c} - \frac{4\pi G_N(p_{c}+\rho_{c})(3p_{c}+\rho_{c})}{3\,{\dd{P}}/{\dd\rho}|_c}r_{i}^{2}
    + \mathcal{O}\left(r_{i}^{4}\right),\label{eq:rhocen}
\end{align}
where $r_i$ is a small radius near the center of the NS which was taken as $0.1$ cm so that $r_i\ll R$, and subscript c's refer to central values.~The central pressure $P_c$ is determined from the EOS once the central density $\rho_{c}$ is specified, and $\alpha_{c}$ is arbitrary so will be set to unity for integration proposes.~For each value of $\rho_c$ (with $\alpha_c=1$), Eqs.~\eqref{eq:dp1} and  \eqref{eq:alpOD1} can be integrated outward with Eq.~\eqref{eq:rhocen} as initial conditions until the density vanishes, $\rho(R) =0$, which defines the radius $R$ of the NS.~Then, Eq.~\eqref{eq:alpOD1} is solved outside by setting $\rho$ and P to zero, with $\alpha(R)$ and $\alpha'(R)$ as boundary conditions.~The mass of the NS is calculated by matching the exterior numerical solution at some distance ${r_b\approx 10^9\text{cm}\gg R}$ to the asymptotic expansion 
\begin{align}
\label{eq:alpa_asymptotic}
    \lim_{r\to\infty}{\alpha(r)} &= A - \frac{{2}G_NM}{r} -\frac{1}{2}\left(\frac{{2}G_NM}{r}\right)^{2} + \mathcal{O}\left(\frac{1}{r^{3}}\right)\\
\end{align}
where A is a integration constant which arises because $\alpha_{c}$ is arbitrary.~Note that this asymptotic expansion has the same form as GR to $\mathcal{O}(r^{-1})$.~Without loss of generality, the constant A can be set to unity since this is just a redefinition of the time coordinate of the form $t\rightarrow e^{-A}t$.~This change of coordinates is equivalent to the transformation $\alpha\rightarrow \alpha - A$, which will be used in the following section to construct slowly rotating NSs.~It is important to note that the expansion in Eq.~\eqref{eq:alpa_asymptotic} is only valid in the intermediate regime where $r \gg R $ but much smaller than the scale where the theory transitions to the MOND regime ($\sim$kpc).

Fig.~\ref{fig:MR_APR} shows the mass-radius relations for different values of $\lambda_{\rm s}$ and $K_{\rm B}$ found by solving the stellar structure equations above.~The APR equation of state \cite{PhysRevC.58.1804} was assumed but other choices give qualitatively similar behavior~\cite{Reyes:2024oha}.~Also shown are X-ray measurements of the pulsars~PSR J0740+6620 \cite{Miller_2021}, PSR J0030+0451 \cite{Riley_2019}, and PSR J0348+0432 \cite{Antoniadis_2013}~These figures demonstrate AeST's compatibility with observations.~Explanations of the effects of varying the parameters on the mass-radius relations can be found in \cite{Reyes:2024oha}.

\subsection{First-Order Equations and Slowly Rotating Neutron Stars}
\begin{figure}[ht]
    \centering
\includegraphics[width=0.9\textwidth]{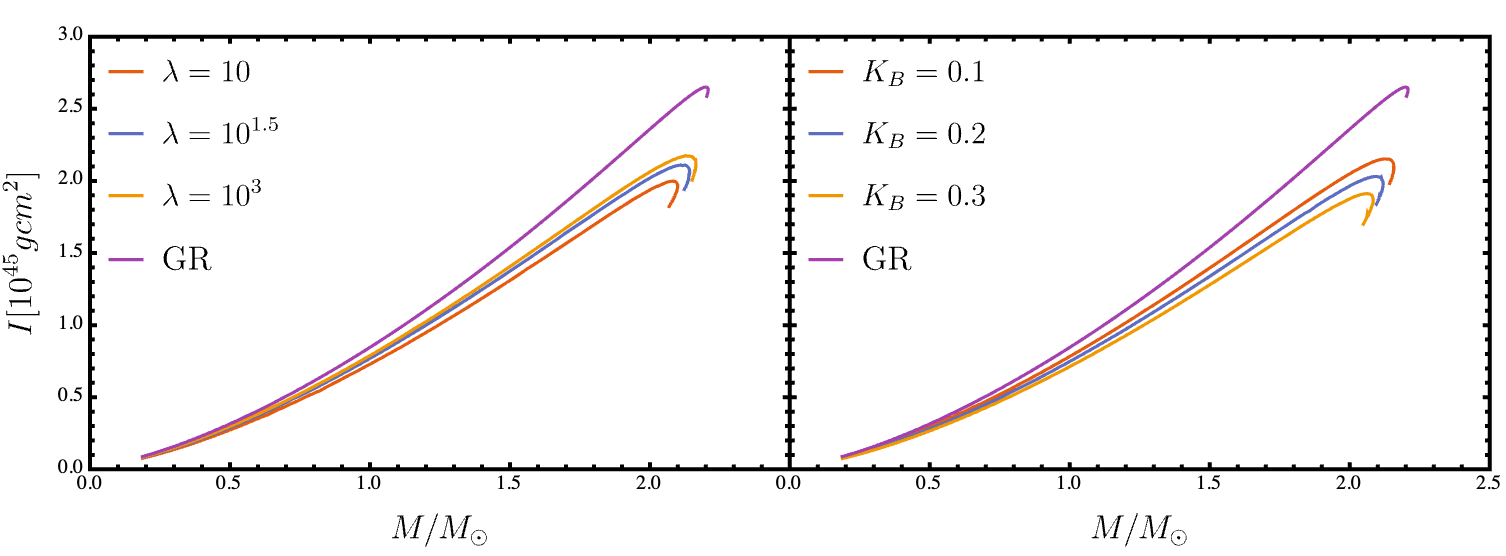}
    \caption{I-M relations for the AeST parameters indicated in the figure assuming the APR EOS.~The left panel shows the effect of varying $\lambda_{s}$ at fixed $K_{B}=0.1$, and the right panel shows the effect of varying $K_{B}$ at fixed $\lambda_{s}=10^2$.}
    \label{fig:MI_APR}
\end{figure}

At first-order in slow rotation, $\mathcal{O}(\varepsilon)$, the only non-vanishing dynamical equations are the $\left(t,\phi\right)$-component of the modified Einstein equations and the $\phi$-component of the vector equation, from which we find 
\begin{equation}
    \omega ''(r)+\frac{\mathcal{W}(r;K_B,\lambda_{s},\alpha)}{8 (\lambda_{\rm s} +1) r \left(\lambda_{\rm s} +1-4 \pi  G_{N} \lambda_{\rm s}  (K_B-2) r^2 P\right)}=0,\label{eq:omegaOD}
\end{equation}
and
\begin{equation}
    B ''(r)+\frac{\mathcal{S}(r;K_B,\lambda_{s},\alpha)}{8 (\lambda_{\rm s} +1) r \left(\lambda_{\rm s} +1-4 \pi  G_{N} \lambda_{\rm s}  (K_B-2) r^2 P\right)}=0\label{eq:BOD}
\end{equation}
where $\mathcal{W}$ and $\mathcal{S}$ are long and cumbersome, so we do show them here.~Their explicit form and a complete derivation of these equations can be found in our supplementary code \cite{Reyes:2025aether}.~As shown in the code, the $\mathcal{O}(\varepsilon)$ scalar-field, $\phi_1(\vec{r})$ is zero.~Equations \eqref{eq:BOD} and \eqref{eq:omegaOD} can be solved near the center of the star up to $\mathcal{O}(r^2)$  to find: 
\begin{align}
    \omega(r_i) &= \omega_{c} -\frac{4 \pi  G_{N} \left( \left(\rho_{c}+3P_c \right)\left(5 e^{\frac{\alpha_c}{2}} B_{0} (\lambda_{s}  K_{B}+2)+6 \lambda_{s}  (K_{B}-2) \omega_{c}\right)\right)}{15 (\lambda_{\rm s} +1)}r_i^{2}
    + \mathcal{O}\left(r_i^{4}\right)\\ 
   B(r_i)&= B_{0}r_i^{2} + \frac{8 \pi  B_{0} G_{N} \left( \rho_{c} ((9 \lambda_{\rm s} +7) K_{B}+4)+3 ((\lambda_{\rm s} -1) K_{\rm B}+4) P_{c}\right)}{5 (\lambda_{\rm s} +1) K_{B}}r_i^{4} +\mathcal{O}\left(r_{i}^{6}\right)
\end{align}
where $\omega_{c}$ and $B_{0}$ are arbitrary constants.~Once $\omega_{c}$ and $B_{0}$ are specified, Eqs.~\eqref{eq:omegaOD} and \eqref{eq:BOD} can be integrated from $r_i$ to the radius of the of the NS, and then further integrated outside the star by  setting $P=\rho=0$ and using $\omega(R)$, $B(R)$, and their respective derivatives as boundary conditions.~Expanding \eqref{eq:omegaOD} and \eqref{eq:BOD} far from the star we obtain the asymptotic behavior: 
\begin{align}
\label{eq:omega_asymptotic}
 \lim_{r\to\infty}{\omega(r)} &= \Omega_{*} - \frac{{2}G_NJ}{r^{3}}   +\frac{2G_Ne^{\frac{\alpha_{c}}{2}}M (\lambda_{s}  K_{B}+2)D}{4 (\lambda_{s} +1)r^{4}}+\mathcal{O}\left(\frac{1}{r^{5}}\right)\\
\lim_{r\to\infty}{B(r)} &= \frac{D}{r} +  \frac{e^{-\frac{\alpha_{c}}{2}} G_{N} M \left(2 e^{{\frac{\alpha_c}{2} }} D ((3 \lambda_{\rm s} +2) K_{B}+2)+6 G_{N} J (\lambda_{\rm s}  K_{B}+2)\right)}{4 (\lambda_{\rm s} +1) K_{B}r^{2}} + \mathcal{O}\left(\frac{1}{r^{3}}\right)
\end{align}
where $J$ is the angular momentum of the NS and $D$ is a integration constant that arises because $B_0$ in \eqref{eq:BOD} is arbitrary.~These expressions are derived in our supplementary code \cite{Reyes:2025aether}.~The leading order term in the $B(r)$ asymptotic expansion scales as $r^{-1}$, but while integrating Eqs.~\eqref{eq:omegaOD} and \eqref{eq:BOD} we found that $B(r)$ can acquire a spurious quadratic behavior at large distances due to a numerical instability.~To understand the origin of this, let us consider Eq.~\eqref{eq:BOD} in Minkowski space.~Setting the metric potentials and matter fields to zero we find
 \begin{align}
B^{\prime\prime}\left(r\right)  -2B\left(r\right)/r^{2} =0
\end{align}
which has the general solution 
\begin{align}
\label{eq:Bsolnumeric}
 B\left(r\right) =a/r+br^{2},
\end{align}
where $a$ and $b$ are integration constants.~The term proportional to $b$ is absent when higher-order terms in the metric are included but as one integrates to larger distances where the space-time is asymptotically flat, numerical errors compound and inevitably induce the quadratic term, which then grows with radial coordinate.~

To remove this anomalous solution, we solve \eqref{eq:omegaOD} and \eqref{eq:BOD} numerically for two sets of initial conditions  $\left(B_{0},\omega_{c}\right) $ and obtain two sets of solutions $\xi_i^{(1)}$ and $\xi_i^{(2)}$, where $\xi_i = \{B(r),\omega(r)\}$.~Since \eqref{eq:omegaOD} and \eqref{eq:BOD} are linear in $\omega$ and $B$, any linear combination of the form $\xi_i$ = $C_{1}\xi_i^{(1)} + C_{2}\xi_i^{(2)}$ where $C_{1}$ and $C_{2}$ are constants is also a solution.~The quadratic behavior can then be removed by setting $C_{1}b_{1} + C_{2}b_{2} = 0$, where $b_{j}$ is the quadratic coefficient of the $\xi_{j}$ numerical solution; $b$ in Eq.~\eqref{eq:Bsolnumeric}.~We determine $b_1$ and $b_2$ numerically by extracting the leading-order behavior for each solution $B_i$ at large distances then, without loss of generality, we set $C_1$ to unity and solve for $C_2$.

Once asymptotic flatness has been imposed, we can proceed to remove the arbitrariness of $\omega_{c}$ and $B_{0}$ by following the scheme described in \cite{Vylet:2023pkp}, which we briefly review here.~We build two linearly independent solutions $Z_i^{(1)}$ and $Z_i^{(2)}$ with $Z_i=\{\xi_1,\xi_2\}$ found using the method described above.~The linear combination  $Z_i$ = $A^{\prime}Z_i^{(1)} + B^{\prime}Z_i^{(2)}$ is then an asymptotically flat solution, where $A^{\prime}$ and $B^{\prime}$ are constants.~These two constants along with $J$ and $D$ are determined by matching the $r \rightarrow \infty$ asymptotic expansion in Eq.~\eqref{eq:alpa_asymptotic} and its derivative to this linear combination at some large radius $r_{b}\approx 10^{9}{\rm cm} \gg R$.~Finally, once $J$ has been calculated the moment of inertia can be obtained from $I = J/\Omega_{*}$.~

\begin{figure}
    \centering
\includegraphics[width=0.7\textwidth]{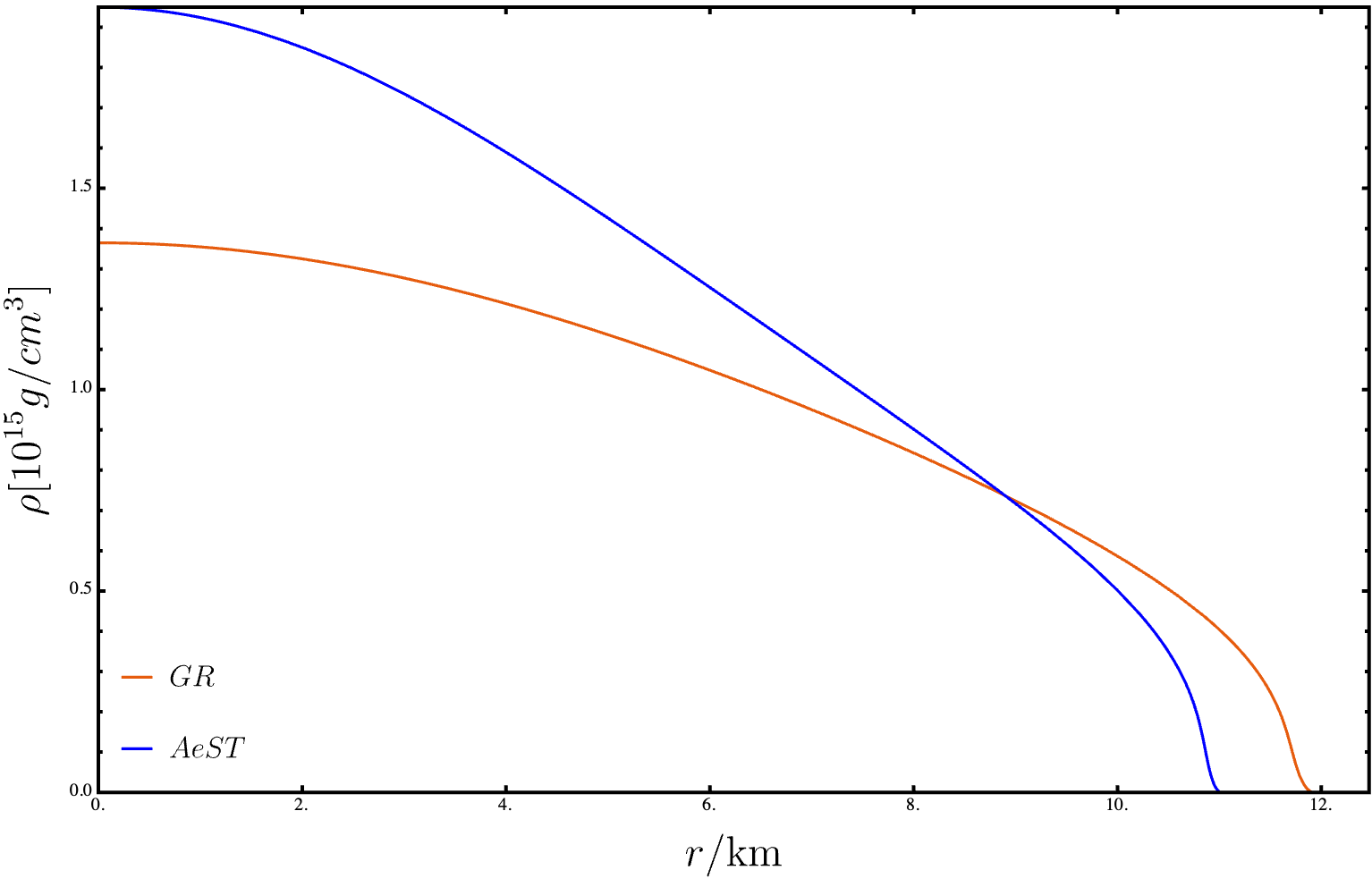}
    \caption{Density profile for a NS with mass of $2.36\msun$ both in GR and AeST with $K_B=0.1$ and $\lambda_{\rm s} = 10^2$.~AeST produces stars with a mass distribution closer to the center of the star.~This behavior is observed across different values of $K_B$ and $\lambda_{\rm s}$.~}
    \label{fig:den}
\end{figure}

Figure \ref{fig:MI_APR} shows the $I$-$M$ relation for different choices of the AeST parameters $\lambda_{\rm s}$ and $K_{B}$ for the APR EOS.~In general, for a fixed NS mass AeST predicts stars with a smaller moment of inertia $I$ than GR.~This can be understood by considering density profiles for neutron stars of fixed mass, and example of which is shown in Figure \ref{fig:den}.~This figure demonstrates that, at fixed total mass, AeST NSs have more mass concentrated towards their centers, implying a smaller moment of inertia.

\begin{figure}[ht]
    \centering
   \includegraphics[width=0.44\textwidth]{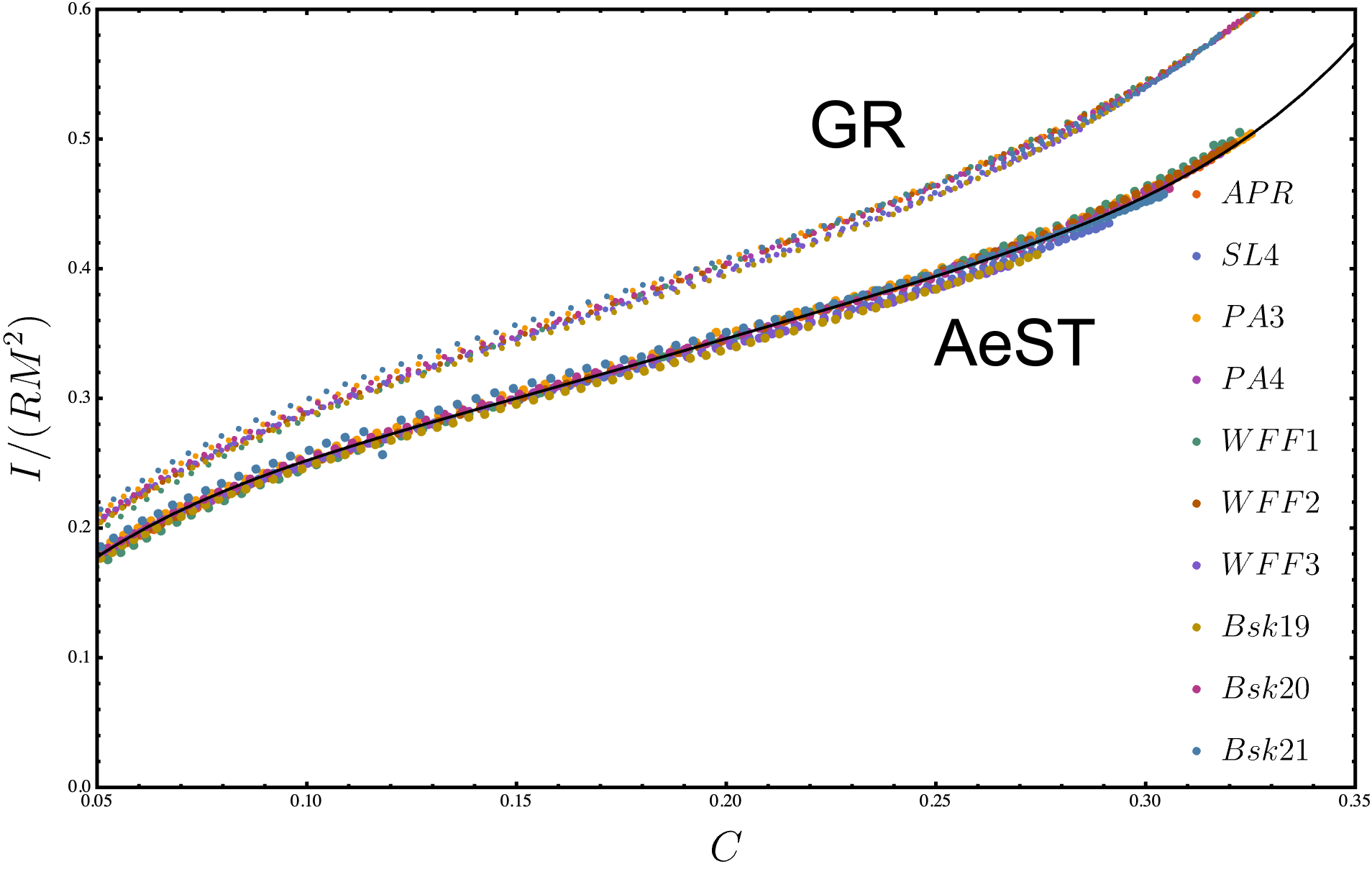}    
\includegraphics[width=0.44\textwidth]{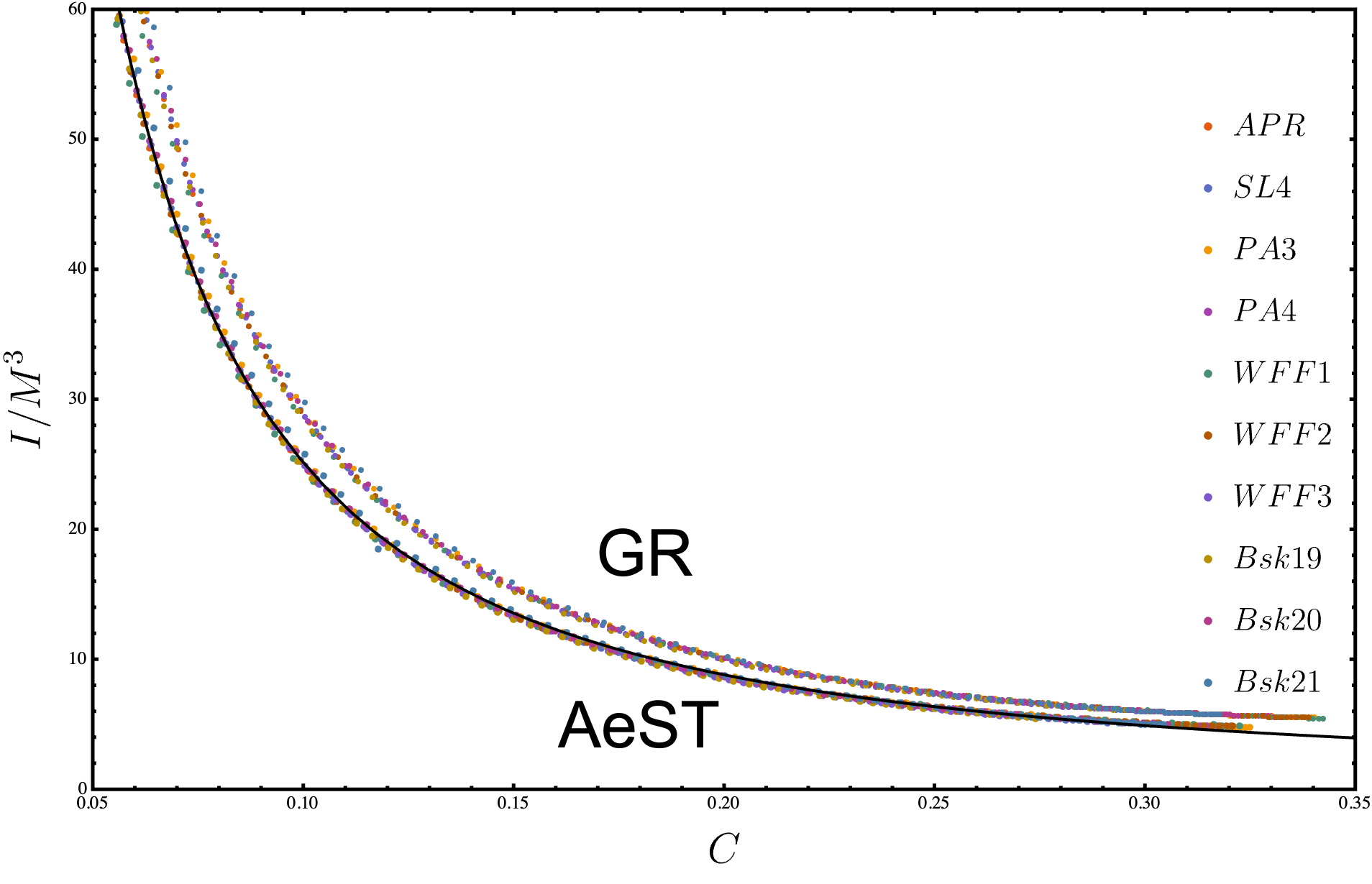}
    \caption{\textbf{Left:}~Dimensionless moment of inertia $I/RM^{2}$ as a function of the compactness, $\cal C$.~The black line shows the fitting function given in \eqref{fit1}.~\textbf{Right:}~Dimensionless moment of inertia $I/M^3$, as a function of the compactness, $\cal C$.The black line here show the fitting function given by \eqref{fit2}.~The colored points in both diagrams correspond to models with a different EOS given in the legend.~The AeST parameters used were $K_{\rm B}=0.1$ and $\lambda_{\rm s} =10$.}
\label{fig:universal}
\end{figure}

It has been demonstrated that GR predicts  approximately universal relations between the dimensionless moments of inertia $\tilde{I}=I/MR^2$ and $\bar{I}=I/M^3$, and the compactness $\mathcal{C}=GM/R$ \cite{Yagi:2013bca,Yagi:2013awa,Breu:2016ufb,Yagi:2016bkt}.~Similar relations are observed in other theories of gravity e.g., \cite{Maselli:2016gxk,Babichev:2016jom,Sakstein:2016oel,Doneva:2017jop,Vylet:2023pkp}.~We calculated $\tilde{I}$, $\bar{I}$, and $\cal C$ in AeST for 10 equations of state indicated in Figure~\ref{fig:universal} for a broad range of $K_{\rm B}$ and $\lambda_{\rm s}$.~We found that the theory predicts approximate universal relations that deviate from the GR predictions in a parameter-dependent manner.~Examples are shown in Fig.~\ref{fig:universal}.

We derived fitting-functions for both $I$--$\cal C$ relations to facilitate  their rapid calculation over a broad range of physically-relevant parameters.~For the $\tilde{I}$-$\cal C$ relation, we found that the function 
\begin{equation}
\tilde{I}_{\rm fit}=a_0(X,K_{B})+a_1(X,K_{B}){\cal C}+a_2(X,K_{B}){\cal C}^{2} + a_3(X,K_{B}){\cal C}^{3} + a_4(X,K_{B}){\cal C}^{4} +a_5(X,K_{B}){\cal C}^{5}
\ ,\label{fit1}\quad
X = \log_{10}(\lambda_{\rm s})
\end{equation}
with coefficients 
\begin{equation}
a_{i}(X,K_{B})=C_{i,0} + \xi_{i}XK_{B} + b_{i}X
+d_{i}K_{B} +g_i\ln(X)+ p_i/X ,\label{eq:fitCoeffs1}
\end{equation}
fits the data well with average relative error $\mid 1 - \tilde{I}/\tilde{I}_{fit}\mid$ across values of $0.1<K_B<0.3$  and $1<\log_{10}(\lambda_{\rm s})<3$ of $1.5\%$ and standard deviation of $0.48\%$.~The coefficients $\{C_{i,0},\xi_i,b_i,d_i\}$ are given in table~\ref{tableI1} and an example of this fit is shown by the black line in Fig.~\ref{fig:universal}.

For the $\bar{I}$-$C$ relation, we find that the function
\begin{equation}
\bar{I}_{\rm fit}=a_1(X,K_{B}){\cal C}^{-1}+a_2(X,K_{B}){\cal C}^{-2}+a_3(X,K_{B}){\cal C}^{-3}
+a_4(X,K_{B}){\cal C}^{-4}\ ,\label{fit2}
X = \ln(\lambda_{\rm s}) 
\end{equation}
with
\begin{equation}
a_{i}(X,K_{B})=C_{i,0} + \xi_{i}XK_{B} + b_{i}X
+d_{i}K_{B} +g_i\ln(X)+ p_i/X ,\label{fitcoef2}
\end{equation} 
provides a good fit with an average relative error of $1.5\%$, and standard deviation of $0.48\%$.~The coefficients $\{C_{i,0},\xi_i,b_i,d_i,g_i,p_i\}$ are given in table~\ref{tableI2} and an example of this fit is again shown by the black line in Fig.~\ref{fig:universal}.

The discovery of the $I$--$C$ relations are the main result of this work.~They may enable future tests of AeST using X-ray and gravitational wave observations along the lines of \cite{Silva:2020acr,Saffer:2021gak} once the higher-order I-Love-Q relations in AeST are derived.~The Neutron star Interior Composition ExploreR (NICER) measured the compactness of a $1.4\msun$ NS to be $C=0.159^{+0.025}_{-0.022}$ \cite{Riley:2019yda,Miller:2019cac,Silva:2020acr}.~Independently, LIGO/Virgo measured the dimensionless tidal deformability $\Lambda=\lambda/M^5$ ($\lambda$ is the Love number) of a $1.4\msun$ NS as $\Lambda=190^{+390}_{-120}$ \cite{LIGOScientific:2018cki}.~Together, these provide a region in the $I$--$\Lambda$ plane that the $I$-Love relation must pass through.~AeST parameters for which this is not the case can be excluded.~Performing this test requires the $I$-$C$ relations we have derived here to convert the NICER measurement of the neutron star compactness to its moment of inertia for use in the $I$-Love relation.



\begin{table}[t]
\centering
\begin{tabular}{ccccccc}
\hline
\hline
Parameter &$a_0$
&$a_{1}$
& $a_{2}$
& $a_{3}$
& $a_4$
& $a_5$

\\
\hline
$C_{i,0}$ &0.0158548 
&6.74109 
&-55.4503
&258.671
&-575.311
&466.988
\\
$\xi_{i}$ & -0.000902438 
& -0.0389251
&-0.16922
&5.31608
&-26.2536
&42.7905
\\
$b_{i}$ & 0.00124156
&-0.0147552
&1.05456
&-11.4717 
&49.0046
&-75.0804
\\
$d_{i}$ & -0.00719414
& -2.42911
&20.7093
&-92.9913
&200.191
&-150.463
\\
$g_{i}$ & -0.00615637 
& -0.383889
& -0.683402
& 31.6818
&-171.747
&293.219 
\\
$p_{i}$ & -0.00996339
& -1.27318
& 5.85208
&10.9717
&-156.378
&329.316
\\
\hline
\hline
\end{tabular}
\caption{Fitting coefficients for the $\tilde{I}-\cal C$ relation in equations \eqref{fit1} and \eqref{eq:fitCoeffs1}.~}
\label{tableI1}
\end{table}

\begin{table}[t]
\centering
\begin{tabular}{ccccccc}
\hline
\hline
{Parameter} &
$a_{1}$
& $a_{2}$
& $a_{3}$
& $a_{4}$

\\
\hline
$C_{i,0}$ &1.05332
&0.271512 
&-0.00467516
&0.000027317
\\
$\xi_{i}$ & 0.0147911
& -0.00502308
&0.0000868944
&$-5.13171\times10^{-7} $

\\
$b_{i}$ & 0.00279832
&0.00215865 
&$6.76488\times10^{-7}$
&$-3.43286\times10^{-7}$
\\
$d_{i}$ & -0.211194 
& -0.108713
& 0.00197305
&-0.0000120605

\\
$g_{i}$ & -0.0501168 
& -0.0294797
& 0.000391721
&$-1.20776\times10^{-6}  $
\\
$p_{i}$ & -0.100108
& 0.0000800608
& 0.00125813
&$-6.7547\times10^{-6}$
\\
\hline
\hline
\end{tabular}
\caption{Fitting coefficients for the $\bar{I}-\cal C$ relation in equations \eqref{fit2} and \eqref{fitcoef2}.~}
\label{tableI2}
\end{table}

\section{Discussion and Conclusions}
\label{sec:conclusions}

Aether Scalar-Tensor theory is a promising alternative to general relativity that can explain all  cosmological observations and resolve the mass discrepancies observed in astrophysical systems from cluster to galactic scales.~Testing it is therefore paramount.~In this work, we derived and solved the equations for slowly rotating neutron stars at first-order, and found that AeST predicts two approximate universal relations between the moment of inertial and compactness ($I$--$\cal C$ relations) that differ from the GR predictions in a parameter-dependent manner.~We constructed fitting functions for both relations that cover a wide range of model space.

These relations are the first step towards probing the theory using X-ray observations of pulsars and gravitational wave observations of binary neutron star mergers.~Both of these systems can provide either measurements of the universal relations between the neutron star mass, compactness, and higher-order mass moments (e.g., Love numbers and quadrupole deformabilities) directly, or provide measurements of neutron star properties at different masses whose consistency with the universal relations can be tested \cite{Silva:2020acr,Saffer:2021gak}.

The most pressing followup investigation is the derivation of the I-Love-Q relations in AeST, which requires extending our derivation to second-order in slow rotation and overcoming potential novel challenges such as the ambiguity in the definition of the quadrupole deformability encountered in GR \cite{Pani:2015hfa,Gralla:2017djj,Creci:2021rkz}.~Another potential study is to extend the range of AeST models that can be probed by generalizing the ansatzes we have made here, either by allowing a radial component for the vector in \eqref{eq:vectorAnt} or by endowing the scalar with time-dependence so that $\phi(t,r)=qt+\varphi(r)$.~Finally, it would be interesting to investigate the behavior of binary pulsars, which have proven to be strongly constraining probes of other theories that include a constrained vector degree of freedom such as Einstein-Aether theory \cite{Yagi:2013ava,Yagi:2013qpa,Gupta:2021vdj} and TeVeS \cite{Kramer:2021jcw}.

\section*{Software}

Mathematica version 12.3.1.0, xCoba version 0.8.6,  xPerm version 1.2.3, xPert 1.0.6 , xTensor version 1.2.0, xTras version 1.4.2.

\section*{Acknowledgments}

We are grateful for discussions with Emanuele Berti, Jo\~{a}o Lu\`{i}s, Constantinos Skordis, and Hector O.~C.~Silva.

\appendix

\clearpage
\bibliography{refs}
\end{document}